\documentclass[superscriptaddress,twocolumn,pre,showpacs,preprintnumbers]{revtex4-1}
\pdfoutput=1
\usepackage{amsmath,amsfonts,amsthm,amssymb}
\usepackage{epsfig,graphics,color,calc,graphicx}
\usepackage[normalem]{ulem}
\usepackage{mathrsfs}
\usepackage{cancel}
\newlength{\pecettawidth}
\begin{document}
%%% Font: commenta la riga che segue se vuoi i font standard
%\fontfamily{ppl}\selectfont
%\logo
\title{Uphill migration in coupled driven particle systems}

\author{Emilio N.M.\ Cirillo}
\email{emilio.cirillo@uniroma1.it}
\affiliation{Dipartimento di Scienze di Base e Applicate per 
             l'Ingegneria, Sapienza Universit\`a di Roma, 
             via A.\ Scarpa 16, I--00161, Roma, Italy.}
%\thanks{ENMC acknowledges Eurandom for the kind hospitality.}

\author{Matteo Colangeli}
\email{matteo.colangeli1@univaq.it}
\affiliation{Dipartimento di Ingegneria e Scienze dell'Informazione e
Matematica,\\
 Universit\`a degli Studi dell'Aquila,
via Vetoio, 67100 L'Aquila, Italy.}

\author{Ronald Dickman}
\email{dickman@fisica.ufmg.br}
\affiliation{Departamento de F\'isica and National Institute 
of Science and Technology for Complex Systems, ICEx, 
Universidade Federal de Minas Gerais, 
C.P.\ 702, 30123--970, Belo Horizonte, Minas Gerais, Brazil.}

%\thanks{The authors thanks....}

\begin{abstract}
In particle systems subject to a nonuniform drive,
particle migration is observed 
from the driven to the non--driven region and vice--versa, depending
on details of the hopping dynamics, leading to apparent violations of
Fick's law and of steady--state thermodynamics.
We propose and discuss a very basic model in the framework of  
independent random walkers on a pair of rings, one of which features biased hopping rates, in which this phenomenon is observed
and fully explained. 
\end{abstract}

%\pacs{64.60.My, 64.60.qe, 05.50.+q, 05.70.Ln, 64.60.an}

\keywords{Driven systems, Random walks, Uphill current}

\preprint{Appunti: \today}

%\ams{prova prova}

\maketitle

\section{Introduction}
\label{s:int}
Particle transport in far--from--equilibrium 
systems often exhibits features that run counter to intuition based on equilibrium thermodynamics.  One class of paradoxical
behavior is that of ``uphill migration,'' that is, migration of particles from regions of lower 
to higher density in the absence of an interaction or external potential, in apparent
violation of Fick's law.  Examples are found in driven lattice gases such as that studied
in \cite{D14}.  In this case, particles with nearest-neighbor excluded-volume interactions
migrate against a density gradient, strengthening rather than reducing it, when half of the
system is subject to a drive, although there is no bias in the hopping rates along the direction
of the density gradient.
The stationary densities in the driven and undriven cannot be predicted by equating the chemical potentials determined from the
uniform systems.  This is consistent with the work of Guioth and Bertin \cite{guioth2018}, who
show that intensive parameters such as chemical potential can be defined if the
transition rates satisfy a factorization condition. The nonuniform density profile that
arises in the vicinity of the boundaries between driven and undriven regions violates
this condition.
While a simplistic argument in terms of effective diagonal barriers is proposed in \cite{D14}
to understand uphill migration, a detailed explanation based on the exclusion interaction and hopping dynamics is lacking. 
Since these observations have eluded explanation, it is
of interest to have simple, analytically solvable examples in which uphill migration arises.
The purpose of this work is to analyze such a system.

A possible interpretation of the remarkable uphill migration phenomenon discussed above 
is that 
a particle typically spends more time in the region of higher density than in that of lower density. 
%Although a simplistic argument in terms of effective diagonal barriers is mentioned in %\cite{D14}, a detailed explanation of this phenomenon in terms of the exclusion interaction %and hopping dynamics is lacking. 
In this work we propose a simple model of {\it independent} random walkers on two rings, in which migration
between driven and nondriven rings is observed and 
fully explained in terms of the typical times spent by particles in 
the two regions. 
The observed mass transport across different regions of the domain shares some interesting features with the uphill diffusion phenomenon discussed in \cite{CDMP16,CDMP17,CGGV18} in the context of lattice gas or spin models,
and with particle migration on a half-driven ladder \cite{ladder07}.
$\;$

The remainder of this work is organized as follows.  
In Sec.~\ref{s:mod} we define the model and present
its stationary solution. Our analytic and simulation results are
discussed in 
Sec.~\ref{s:mig}, illustrating uphill migration.  
The connection between migration and
particle sojourn times, and a connection between the latter and the gambler's ruin problem, are analyzed in Sec.~\ref{s:soj}.  
Some interesting aspects of the stationary density profiles and currents are considered 
in Sec.~\ref{s:pro}, followed, in Sec.~\ref{s:con} 
by a summary and discussion of our findings.  

\newpage
\begin{figure}[htb]
%\vspace{-12cm}
\begin{picture}(80,180)(180,0)
\centering
\hspace{4 cm}
\includegraphics[width=0.4\textwidth]{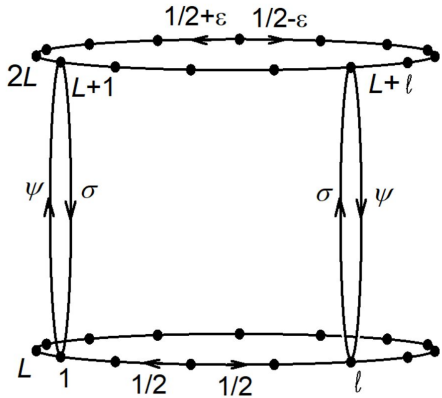} 
\end{picture}
%\vspace{-6.5cm}
\caption{Schematic representation of the model.}
\label{fig:ccd000}
\end{figure}

%\begin{figure}[htb]
%\vspace{-6cm}
%\centering
%\includegraphics[width=0.85\textwidth]{fig-ccd-000.pdf} 
%\vspace{-6.5cm}
%\caption{Schematic representation of the model.}
%\label{fig:ccd000}
%\end{figure}

\section{Model and stationary solution}
\label{s:mod}

We define a particle system of independent random walkers on the 
lattice $\Lambda=\{1,\dots,2L\}$.
The \emph{lower ring} consists of sites $1,...,L$ and the
\emph{upper ring} of sites $L+1,...,2L$, each with periodic boundaries. 
Mass exchange between the rings is allowed via two \emph{channels},
consisting of two fixed pairs
of sites: $(1,L+1)$ and $(\ell,L+\ell)$, 
with $2\le\ell\le L$ (see figure~\ref{fig:ccd000}). 
The distance between channels is $\Delta=\ell-1$.
The number of particles at site $i$ is denoted by $n_i$ and
the total number of particles by $N$.

Each particle performs a random walk on the lattice with \emph{jump 
rate} from site $i$ to site $j$ denoted by $W_{j,i}$ and defined as follows
\begin{equation}
\label{ccd000}
W_{i\pm1,i}=\frac{1}{2}
\;\textrm{ and }\;
W_{L-1,L}=W_{1,L}=\frac{1}{2}
\end{equation}
for $i=2,\dots,\ell-1,\ell+1,\dots,L-1$,
\begin{equation}
\label{ccd010}
W_{L,1}=W_{2,1}=\frac{1}{2}
\;\textrm{ and }\;
W_{L+1,1}=\psi
\end{equation}
with $\psi\ge0$, 
\begin{equation}
\label{ccd020}
W_{\ell\pm1,\ell}=\frac{1}{2}
\;\textrm{ and }\;
W_{L+\ell,\ell}=\sigma
\end{equation}
with $\sigma\ge0$.
The jump rates 
starting 
from a site in the upper ring are chosen similarly, the 
main difference being the presence of the \emph{drift}
$\varepsilon\in[0,1/2]$:
\begin{equation}
\label{ccd030}
W_{i\pm1,i}=\frac{1}{2}\pm\varepsilon,\,
W_{2L-1,2L}=\frac{1}{2}-\varepsilon,\,
W_{L+1,2L}=\frac{1}{2}+\varepsilon
\end{equation}
for $i=L+2,\dots,L+\ell-1,L+\ell+1,\dots,2L-1$,
\begin{displaymath}
\label{ccd040}
W_{2L,L+1}=\frac{1}{2}-\varepsilon,
W_{L+2,L+1}=\frac{1}{2}+\varepsilon,
W_{1,L+1}=\sigma
\end{displaymath}
and
\begin{displaymath}
\label{ccd050}
W_{L+\ell\pm1,L+\ell}=\frac{1}{2}\pm\varepsilon
\;\textrm{ and }\;
W_{\ell,L+\ell}=\psi.
\end{displaymath}
In figure~\ref{fig:ccd000}
we show a schematic representation of 
the model with the particle hopping rates. 
The model can be recast in the 
the language of the 
zero range process \cite{EH05} with site intensity proportional 
to the number of particles occupying the site.

For any site $i$ we let
\begin{equation}
\label{ccd400}
W_i
=
\sum_{j\neq i}W_{j,i}
\end{equation}
be the total rate at which a particle leaves site $i$. 
Note that $W_i=1$ for any $i$, except for sites $1$, $\ell$,
$L+1$, and $L+\ell$ delimiting the channels.
We also let 
\begin{equation}
\label{ccd405}
W_{i,i}
=-W_i.
\end{equation}

For a given walker, 
let $\pi_i$ be the steady--state probability that it occupies 
site $i$. By standard results on Markov Processes, see, i.e., 
\cite[Definition~2.3 and Exercise~2.42]{Ligget},
we have that for any $j$
\begin{equation}
\label{ccd410}
\sum_{i}\pi_iW_{j,i}=0,
\end{equation}
which can be rewritten as 
\begin{equation}
\label{ccd420}
\pi_j
=
\sum_{i\neq j}\pi_i\frac{W_{j,i}}{W_j}
.
\end{equation}

Exploiting independence, 
in the stationary state the mean number of particles at the generic 
site $i$ is 
\begin{equation}
\label{ccd470}
\rho_i
=
N\pi_i
.
\end{equation}

The main quantities of interest in this note are the lower 
and upper density profiles $\rho_i$ with $i=1,\dots,L$ and 
$i=L+1,\dots,2L$, respectively, and the average \emph{mass displacement}
between the lower and the upper rings, 
\begin{equation}
\label{ccd480}
\chi
=
\frac{1}{N}\Big[\sum_{i=1}^L\rho_i-\sum_{i=L+1}^{2L}\rho_i\Big]
=
\frac{1}{N}\sum_{i=1}^L(\rho_i-\rho_{L+i})
.
\end{equation}

\begin{figure}
\centering
\includegraphics[width = 0.5\textwidth]{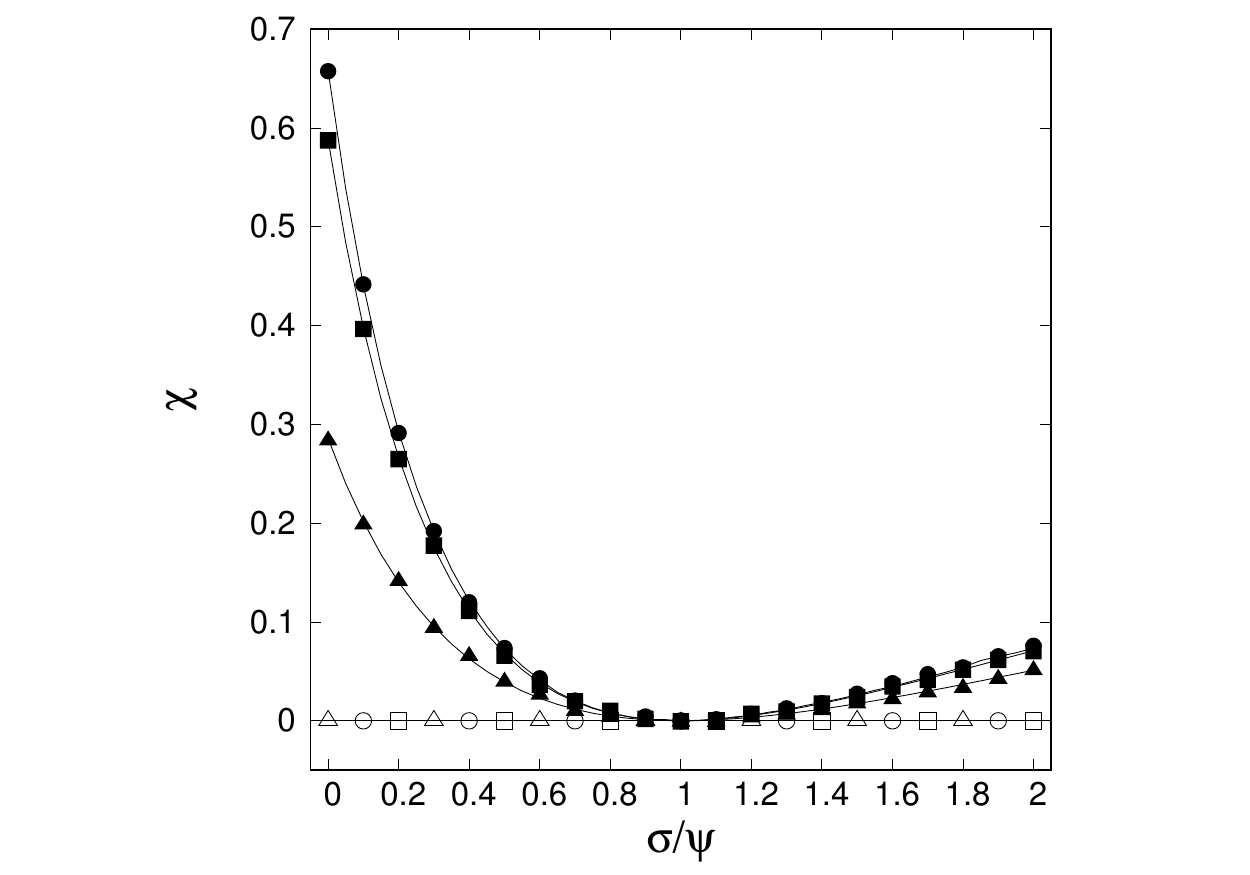} 
\caption{Points: simulation results for mass displacement $\chi$ vs $\sigma/\psi$ for 
$\psi=0.1$ (triangles), $\psi=1$ (squares), and $\psi=10$ (circles), with $L=50$, $\Delta=25$, $N=200$, $\varepsilon=0.1$ (solid symbols),
and $\varepsilon=0$ (open symbols).
Solid lines denote the exact solution.}
\label{fig:fig5}
\end{figure}

\section{Results}
\label{s:res}

In this paper we focus on the behavior of the profiles $\rho_i$ and 
the mass displacement $\chi$ 
as a function of the model parameters. 
For given parameters, we obtain the exact value by first solving
equations \eqref{ccd420} to compute the $\pi_i$'s, and 
then equation \eqref{ccd470} and \eqref{ccd480}. 
In the simulations, the total number of 
particles is $N=200$ and $L=50$.

\subsection{Mass displacement: uphill currents}
\label{s:mig}

\begin{figure}
\centering
\includegraphics[width = 0.5\textwidth]{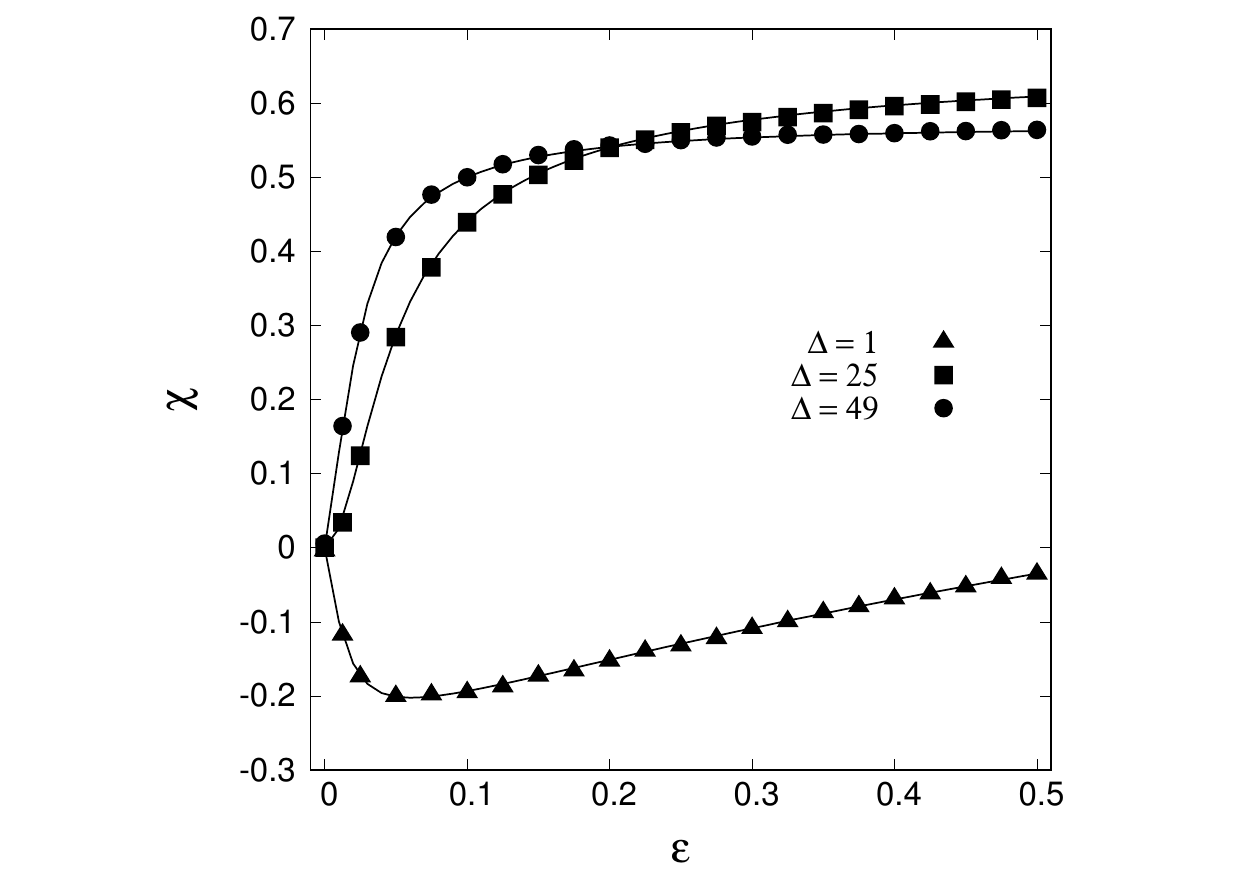} 
\caption{Simulation results for $\chi$ vs $\varepsilon$ 
for $\Delta=1$ (triangles), 
$\Delta=25$ (squares), and $\Delta=L-1$ (circles), with $L=50$,  
$\sigma=10$, $\psi=1$, and 
$N=200$.
Solid lines denote the exact solution.}
\label{fig:fig2}
\end{figure}

\begin{figure}
\centering
\includegraphics[width = 0.5\textwidth]{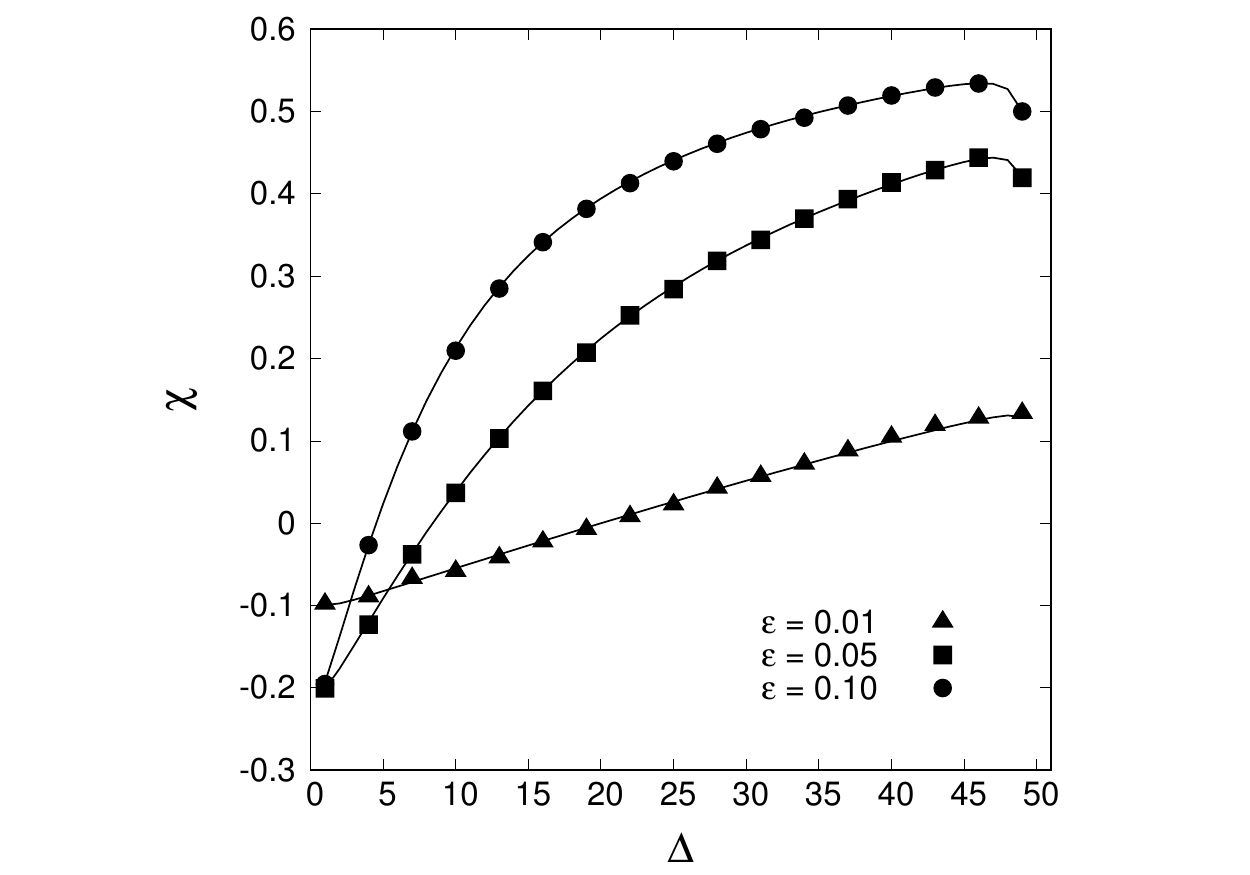} 
\caption{$\chi$ vs $\Delta$ for $\varepsilon=0.01$ (triangles), $\varepsilon=0.05$ (squares),
and $\varepsilon=0.1$ (circles); other parameters as
in figure~\ref{fig:fig2}. Solid 
lines denote the exact solution.}
\label{fig:fig3}
\end{figure}

In this section our principal focus is on the behavior of $\chi$ 
as a function of the model parameters. 
Figs.~\ref{fig:fig5}--\ref{fig:fig3}
compare exact results for $\chi$ with those 
of Monte Carlo simulations.

In figure~\ref{fig:fig5} the mass displacement 
$\chi$ is plotted as a function of
the ratio $\psi/\sigma$ for different choices of the other parameters.
For zero drift, the model is symmetric for any choice 
of parameters
and the mass displacement $\chi$ vanishes.
For $\varepsilon>0$, mass migration is observed ($\chi \neq 0$),
except for the symmetric case, $\sigma/\psi=1$. 
Moreover, in the symmetric case $\sigma/\psi=1$, for any $\varepsilon\ge0$,
the upper and lower profiles $\rho_i$
are uniform, independent of $i$.
The result for $\psi = \sigma$ is expected: in this case symmetry 
is ensured by the fact that in a state in which the mean number 
of particles is the same at each site, the rates at which
particles move upward and downward are equal in each channel.

For various choices of the parameters we find $\chi>0$, namely, 
we observe mass migration towards the non--driven, lower ring.
This phenomenon is not trivial, since the model is designed in 
such a way that upward and downward rates in the two channels
should compensate each other even when $\psi\neq\sigma$. 

In Figs.~\ref{fig:fig2} and \ref{fig:fig3} the mass displacement 
is plotted as a function of the drift $\varepsilon$ and the intra--channel distance $\Delta$,
respectively. These graphs show that 
both positive and negative mass displacements are possible. More precisely, 
given $\varepsilon$ not too large, if the intra--channel distance 
$\Delta$ is small enough then the mass displacement is negative, 
i.e., the mass in the driven ring is larger than 
that in the non--driven one. 
The fact that mass migration is observed in both 
directions is similar to the phenomenon described in \cite{D14}, although in
that work it is connected to the possible jumps that particles may 
perform, whereas here it depends on simple 
geometric features. 

\begin{figure}
\centering
\includegraphics[width = 0.5\textwidth]{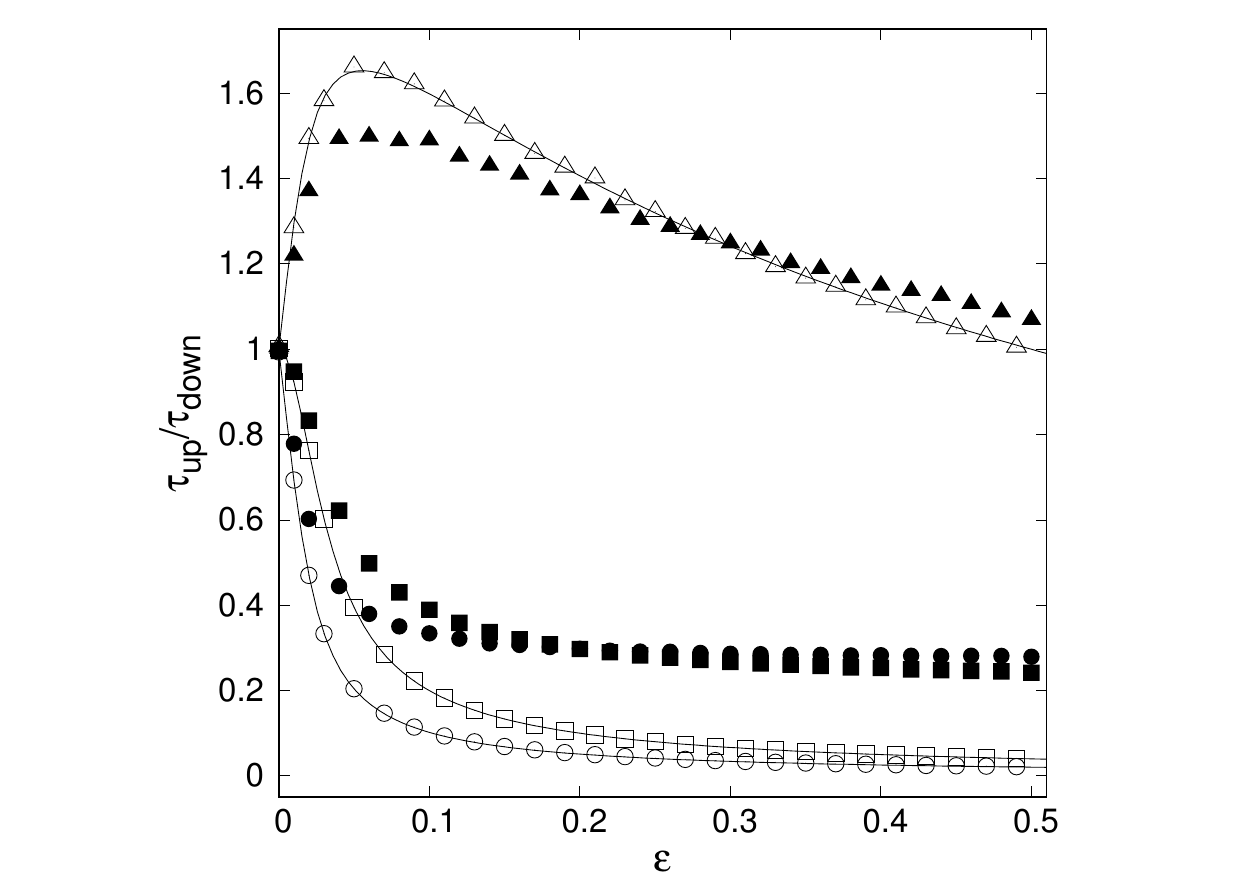} 
\caption{Simulation results for the ratio of the average sojourn
times in the upper and in the lower ring of a single target particle 
as a function of $\varepsilon$, for 
$\Delta=1$ (triangles), $\Delta=25$ (squares), and $\Delta=49$ (circles), 
with $L=50$,  
$N=200$,
$\sigma=10$ and $\psi=1$ (solid symbols),
$\sigma=10^3$ and $\psi=0$ (open symbols).
The solid lines are computed using the gambler's ruin 
expressions for $\tau_\textrm{up}$ 
and $\tau_\textrm{down}$.}
\label{fig:fig4}
\end{figure}

\subsection{Sojourn times and gambler's ruin}
\label{s:soj}
Mass migration can be interpreted in terms of the 
average residence times, called \emph{upper} and \emph{lower sojourn time} and 
denoted respectively by 
$\tau_\textrm{up}$ and $\tau_\textrm{down}$,
that a particle spends in the upper and in the lower ring. 
These times can be defined by attaching a label to each particle $i$,
and following its position over time.  Then $\tau_\textrm{up}$ is defined
as the mean time (over the evolution and over particles) between entering and exiting
the upper ring, in the stationary state; $\tau_\textrm{down}$ is defined analogously.
Sojourn times are similar to 
the residence times that have been 
widely studied for the simple exclusion process
\cite{CKMSpre2016} and
for the simple symmetric random walk 
\cite{CCSpre2018}.
%, and for the linearized Boltzmann equation  \cite{CCkrm2018}.
The main difference is that in those studies 
the geometry of the strip was 
considered and the residence time was defined conditioning the particle 
to exit the strip through the side opposite the one where it started the 
walk. 

The idea is very simple: if the upper sojourn time is larger than 
the lower one, then particles will spend more time in the 
upper ring and, at stationarity, a negative mass displacement will 
be observed. This idea is confirmed by the simulation results plotted 
in figure~\ref{fig:fig4}, where the ratio of the upper and the lower 
sojourn times is plotted as a function of the drift. 
Indeed, comparing data in figure~\ref{fig:fig4} to those in 
figure~\ref{fig:fig2}, a perfect mirror behavior is seen, i.e., 
$\tau_\textrm{up} > \tau_\textrm{down}$ corresponds to 
$\chi < 0$ and vice-versa.
Note, in particular, that the curves for 
$\Delta=25$ and $\Delta=49$ intersect at the same value of 
the drift $\varepsilon$ in both graphs. 

The interpretation in terms of sojourn times also provides a nice explanation of the 
dependence of the sign of $\chi$ on the inter--channel distance. 
In Figs.~\ref{fig:fig2}--\ref{fig:fig4} we consider
$\sigma=10\gg(1/2\pm\varepsilon)$ and 
$\psi=1\approx(1/2\pm\varepsilon)$.
Under this assumption 
particles typically jump from the upper ring to the lower 
one at site $L+1$, and from the lower one to the 
upper one at site $\ell$.

\begin{figure}
\includegraphics[width = 0.3\textwidth]{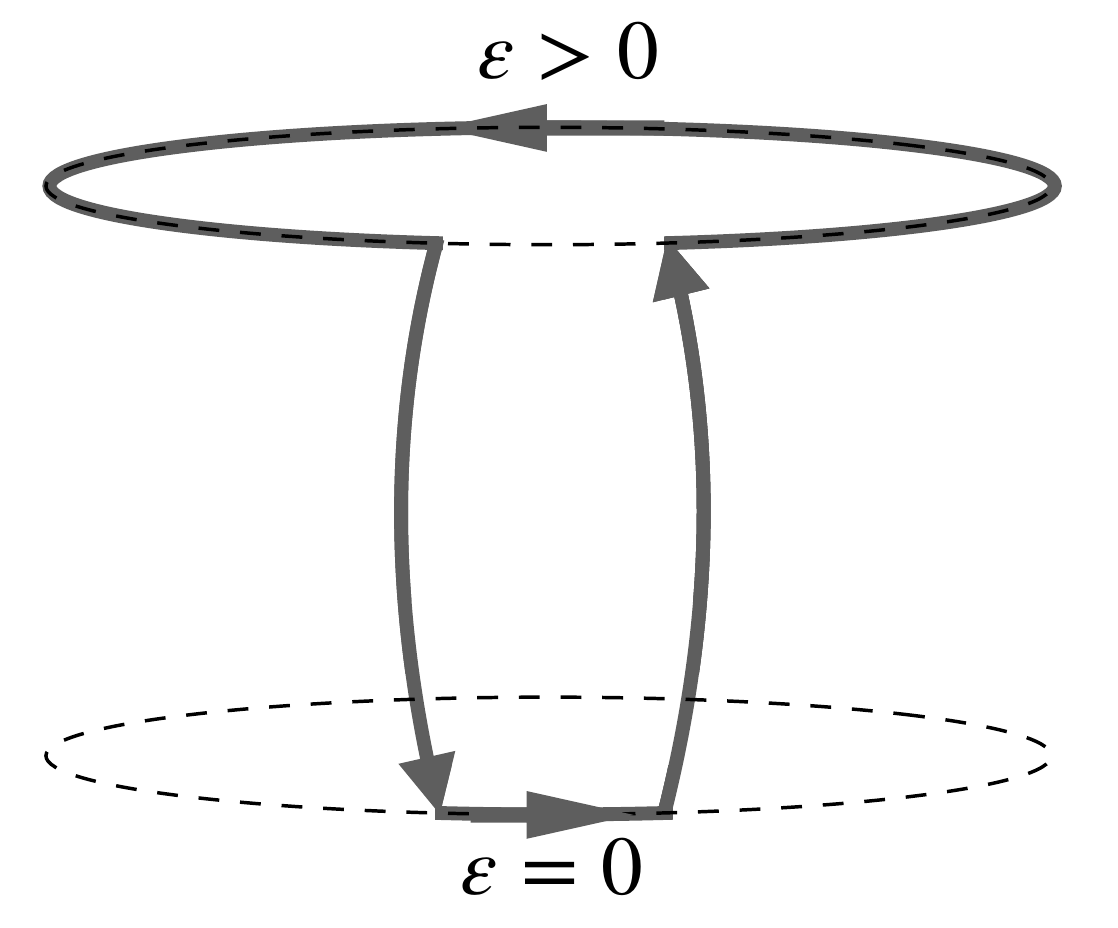} 
\caption{Typical path covered by a particle in the limiting case with $\sigma\gg \psi$ and $\Delta$ small.}
\label{fig:fig8}
\end{figure}

\begin{figure*}
\includegraphics[width = 0.32\textwidth]{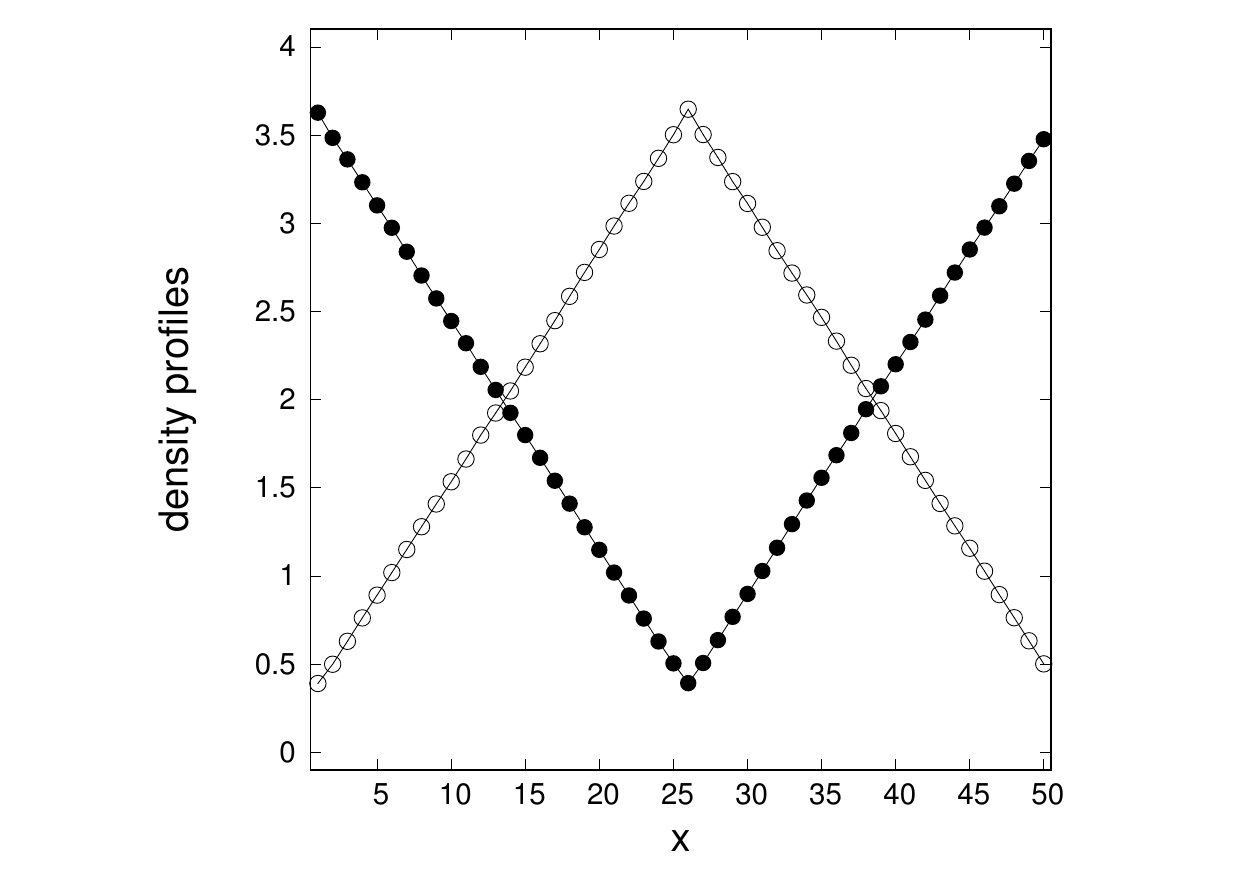} 
\includegraphics[width = 0.32\textwidth]{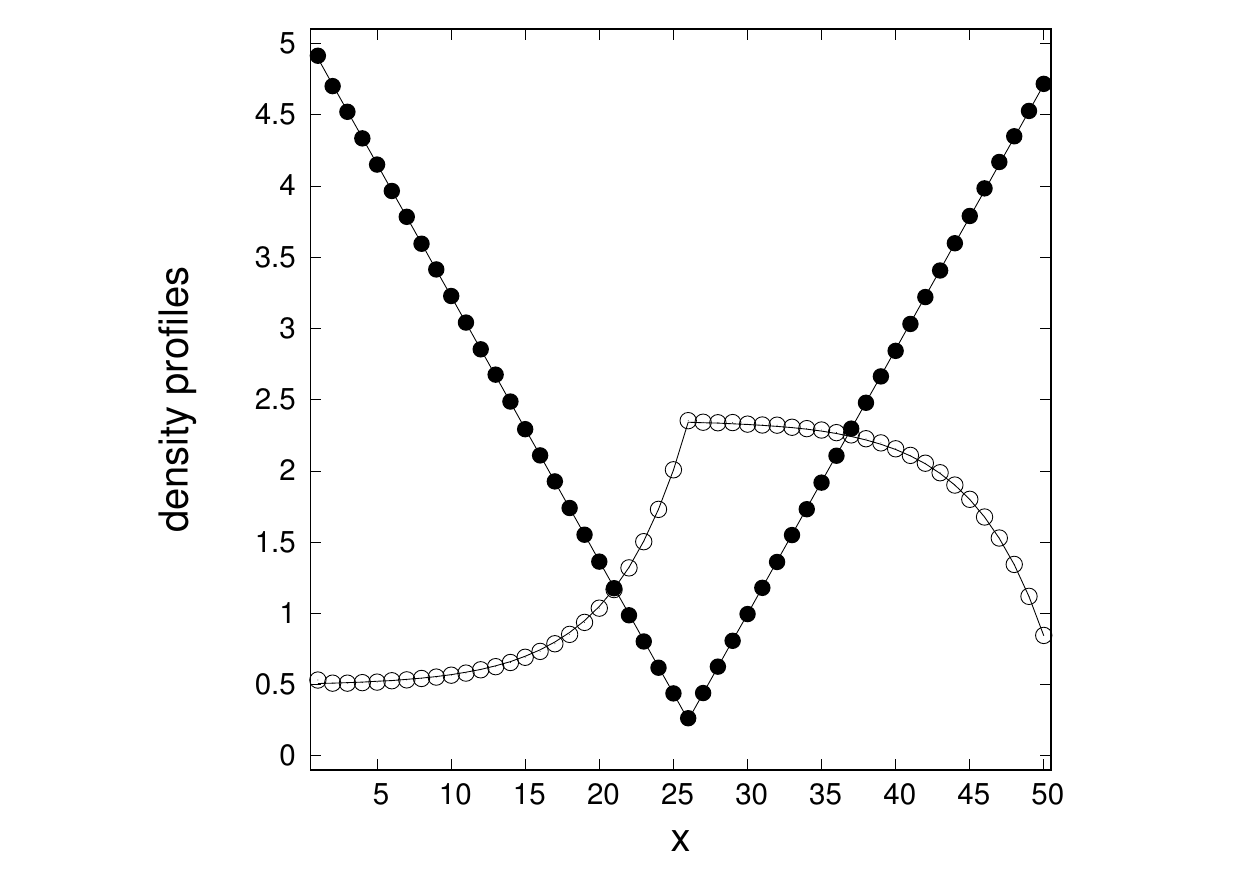} 
\includegraphics[width = 0.32\textwidth]{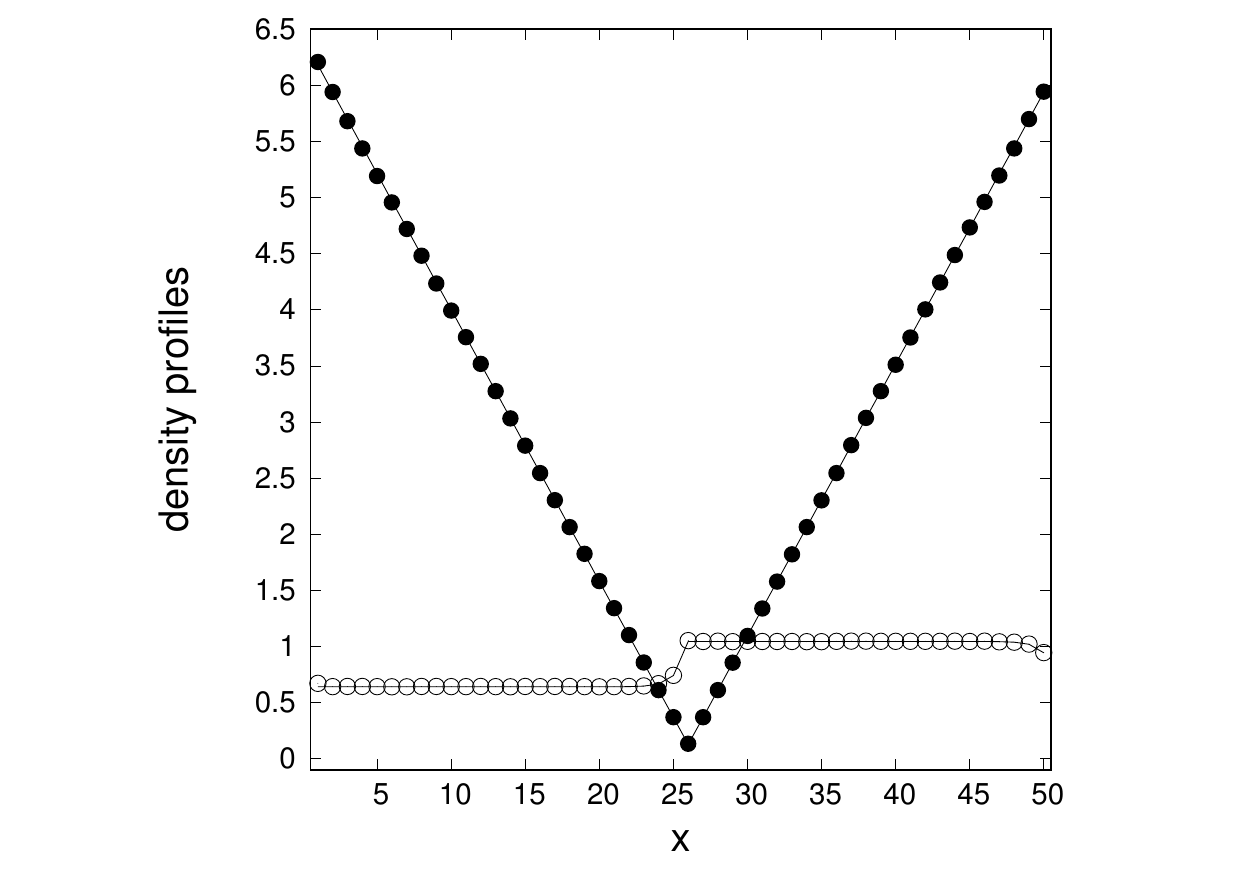} 
\caption{Stationary density profiles for $\sigma=10$ and $\psi=1$, 
with $L=50$, $\Delta=L/2$, and drift 
$\varepsilon=0$ (which corresponds to $\chi=0$), $0.05$ (center, with $\chi=0.2874$), $0.3$ (right, with $\chi=0.5768$). Empty and solid symbols denote, respectively, the density profiles in the upper and the lower rings.
Solid lines denote the exact solution.}
\label{fig:fig6}
\end{figure*}

\begin{figure*}
\includegraphics[width = 0.3\textwidth]{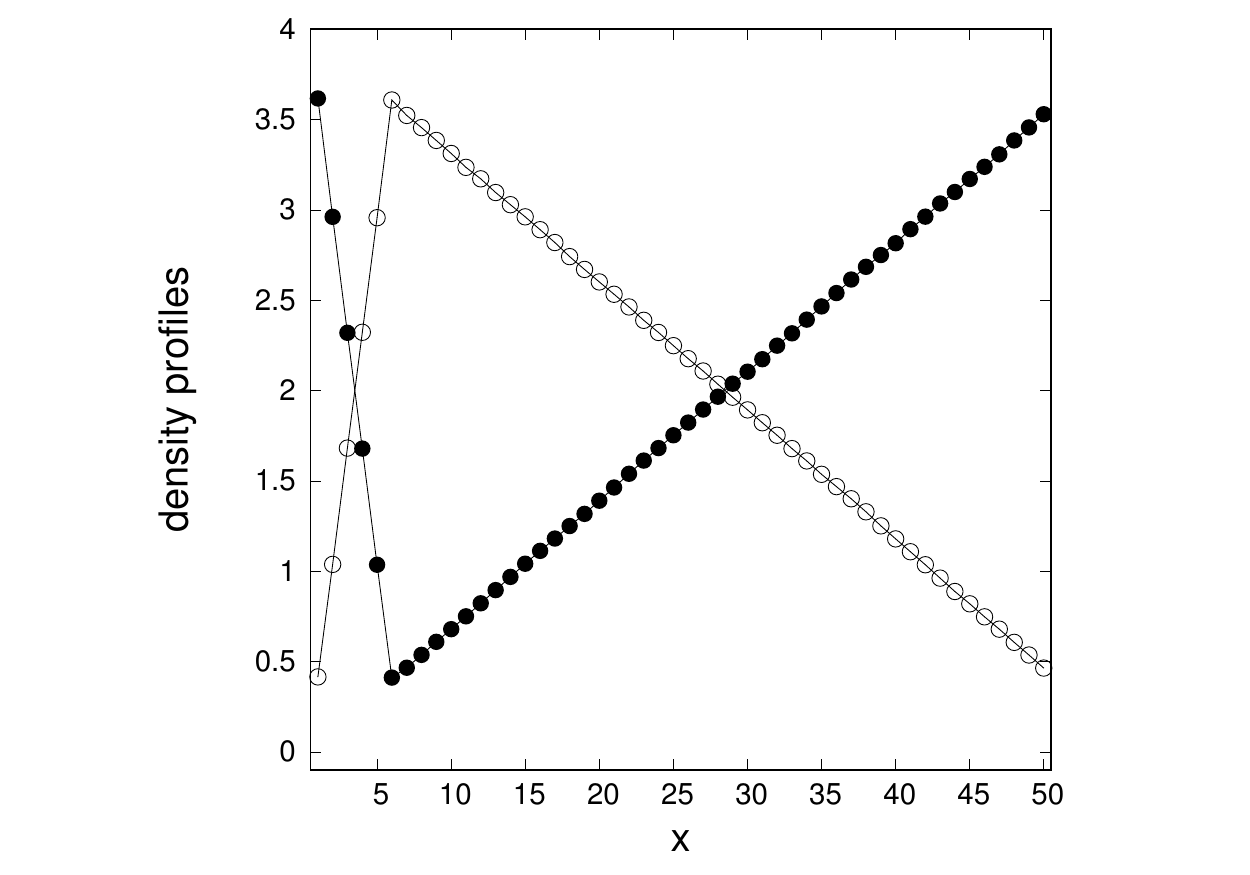} 
\hspace{-1.5 cm}
\includegraphics[width = 0.3\textwidth]{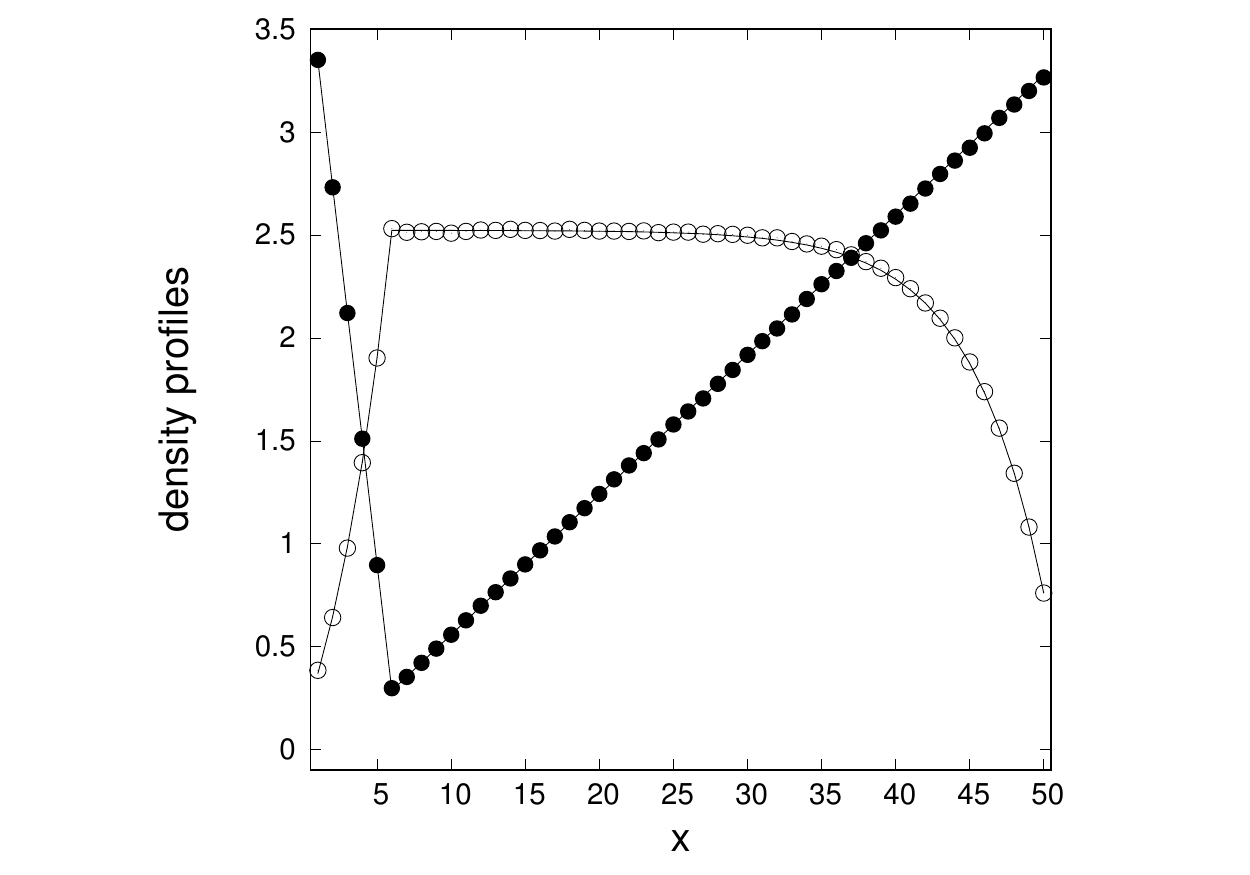} 
\hspace{-1.5 cm}
\includegraphics[width = 0.3\textwidth]{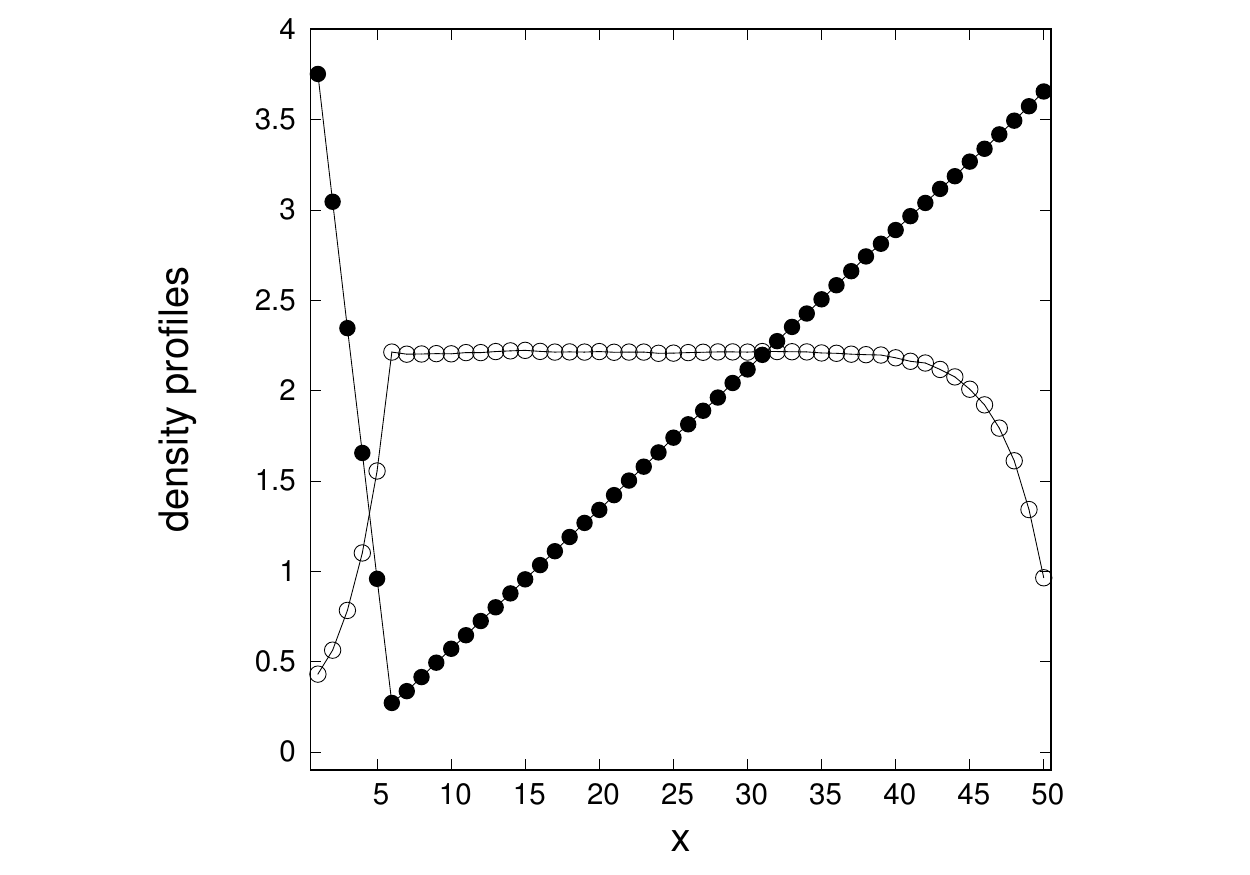} 
\hspace{-1.5 cm}
\includegraphics[width = 0.3\textwidth]{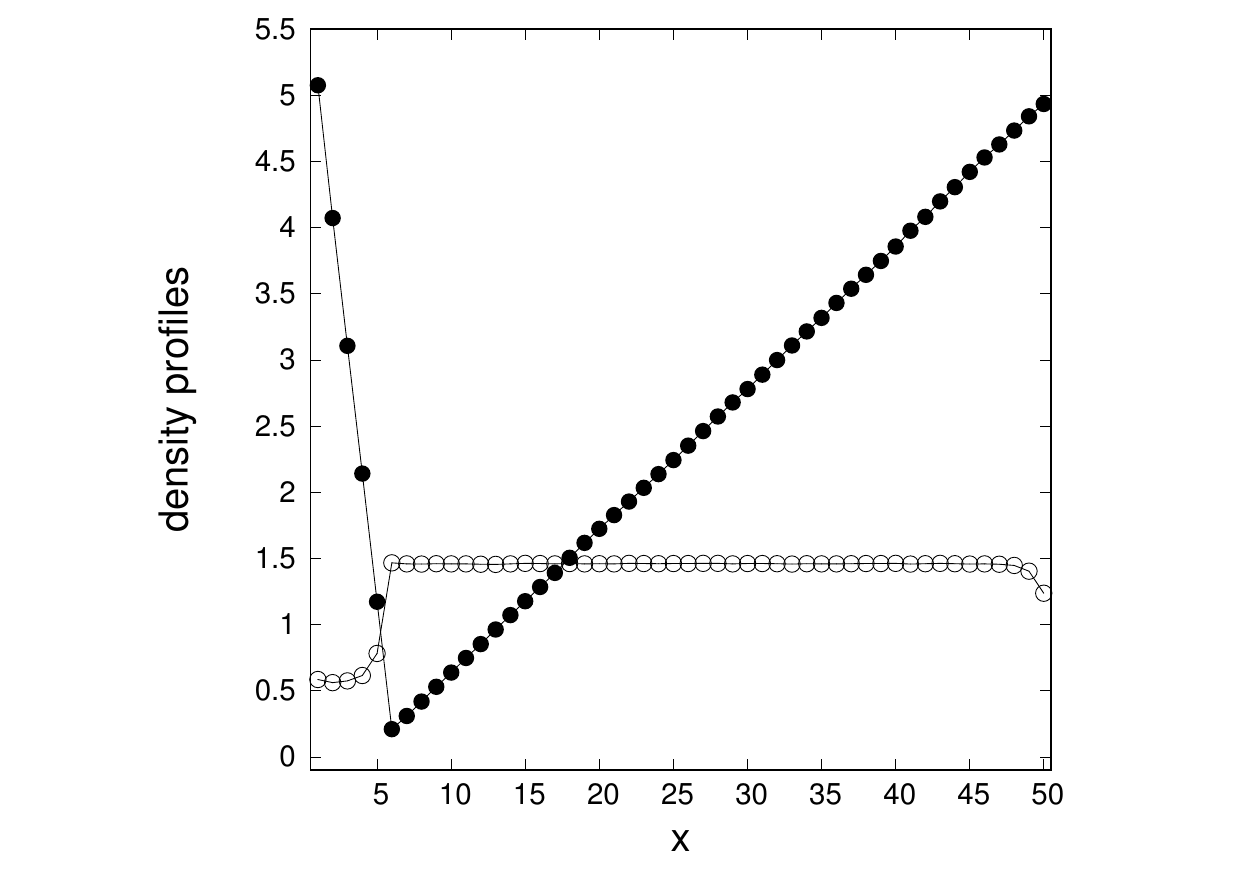} 
\caption{Stationary density profiles for $\sigma=10$ and $\psi=1$, 
with $L=50$, $\Delta=5$, and, from the left to the right, drift 
$\varepsilon=0$ (which corresponds to $\chi=0$), $0.05$ (with $\chi= -0.0929$), $0.0897$ (with $\chi= 0.0002$), and $0.3$ (with $\chi= 0.3145$). Empty and solid symbols denote, respectively, the density profiles in the upper and the lower rings.
Solid lines denote the exact solution.}
\label{fig:fig7}
\end{figure*}

Hence, the upper sojourn time is quite close to the typical time
that a particle needs to move from site $L+\ell$ to site $L+1$ along the 
upper ring, whereas
the lower sojourn time is quite close to the typical time
that a particle needs to move from site $1$ to site $\ell$ along the 
lower ring.
In view of this remark, due to the presence of the drift, the upper 
sojourn time is, for most choices of the parameters, 
smaller than the lower one, explaining why, in general the mass migration is observed 
towards the lower, non--driven ring. The contrary is observed when $\Delta$ 
is small.  Indeed, in this case the lower sojourn time is substantially 
smaller than the upper one. Although no drift is present, the lower 
sojourn time is smaller than the upper one 
because the distance to be covered is much smaller than in the upper, driven ring, cf. 
figure~\ref{fig:fig8}.

This latter observation suggests a way of estimating the sojourn times that 
should be quite accurate for $\sigma\gg 1\gg\psi$. 
Indeed, in this case one can use a classic result of 
probability theory, namely, the gambler's 
ruin estimate of the duration of the game, see 
\cite[Chapter~XIV, Section~3]{Feller1968} 
and 
\cite{CCpreprint2018}, to obtain
\begin{equation}
\tau_\textrm{down}
\simeq
(\ell-1)(L-\ell+1)
\end{equation}
and
\begin{equation}
\tau_\textrm{up}
\simeq
\frac{\ell-1}{1-2\Big(\frac{1}{2}+\varepsilon\Big)}
-
\frac{L}{1-2\Big(\frac{1}{2}+\varepsilon\Big)}
\frac{1-\Big(\frac{\frac{1}{2}-\varepsilon}{\frac{1}{2}+\varepsilon}
      \Big)^{\ell-1}}
     {1-\Big(\frac{\frac{1}{2}-\varepsilon}{\frac{1}{2}+\varepsilon}
      \Big)^{L-1}}
.
\end{equation}
The ratio $\tau_\textrm{up}/\tau_\textrm{down}$ computed using the 
above expressions is plotted in figure~\ref{fig:fig4}, where, as expected, 
the agreement is poor for $\sigma=10$ and $\psi=1$, 
while it is strikingly perfect for $\sigma=10^3$ and $\psi=0$.
Indeed, in such an extreme case, particles move along the two rings 
closely following the gambler's ruin rules. The sole difference with respect 
to the gambler's ruin problem is in the tiny 
probability $1-\sigma/(1+\sigma)$ that a particle at site $L+1$ 
does not jump to the site $1$ and a particle at the site 
$\ell$ does not jump to $L+\ell$. 

\subsection{Density profiles and steady--state currents}
\label{s:pro}

This section is devoted to a brief discussion of the 
typical stationary particle density profiles, which depend on both the drift $\varepsilon$ and 
on the ratio $\sigma/\psi$.
As explained above, 
we perform a series of MC simulations to investigate the stationary 
density profiles in the two rings. The initial datum, in our simulations, 
is represented by uniform density profiles, with the same number of particles 
shared by the two rings. The numerical results are then compared 
to the exact ones obtained as explained at the beginning of
Section~\ref{s:res}.

We have already mentioned that
in the symmetric case $\sigma/\psi=1$, for any $\varepsilon\ge0$,
the upper and lower profiles $\rho_i$
are uniform, namely they do not depend on the site $i$, and $\chi=0$.
This is expected since, in such a case, 
in a state in which the average number 
of particles is the same at each site, the rates at which
particles move upward and downward are equal in each channel.
 
Setting $\psi\neq\sigma$ yields nonuniform stationary profiles in the 
two rings. In particular, figure~\ref{fig:fig6}, for 
$\varepsilon=0$, shows the onset of piecewise 
linear density profiles (i.e., stationary solutions to the discrete
Laplace equation) in both rings, with 
kinks at the locations of the two channels.  For reasons of  
symmetry, there is no net mass transfer between the rings, hence 
we again have $\chi=0$.

The center and the left panel in figure~\ref{fig:fig6} show 
how the profile shape is modified when the drift in the upper channel 
is different from zero, which is the paradigmatic case that leads
to ``uphill currents'', namely states with $\chi \neq 0$. 
We notice, indeed, that, since $\chi>0$ holds in the cases shown in the figure, the overall mass residing in the upper ring, 
in the stationary state, is less than that in the lower ring.
Moreover, it also worth noting that the profiles are not 
linear in the upper ring due to the 
presence of the drift.

The plots in figure~\ref{fig:fig7} refer to an 
inter--channel distance $\Delta=5$, small enough to produce a 
non--monotonic behavior of $\chi$ as a function of $\varepsilon$, already highlighted in 
figure~\ref{fig:fig2} for $\Delta=1$.
In particular, for small values of the drift, particles gather on the upper ring ($\chi<0$), whereas for larger values of the drift they tend to move downwards to the lower ring ($\chi>0$). 

We have also learned that in our model there are two  classes of stationary states showing no uphill diffusion, as also visible in figure~\ref{fig:fig5}. The first class is obtained by setting $\psi=\sigma$ and/or $\varepsilon=0$: in particular, the case with $\psi=\sigma$ and 
$\varepsilon=0$ corresponds to a state with vanishing current and whose density profiles are uniform (except for the coupling sites) over the two rings. The second class includes the states for which the (moderate) effect of the drift in the upper ring is exactly compensated by the (small) inter--channel distance covered by particles in the lower ring, see the third panel from the left in 
figure~\ref{fig:fig7}. These are nonequilibrium steady states characterized 
by nonuniform density profiles and nonvanishing currents, which nevertheless produce 
no net mass transfer between the rings.

\begin{figure*}
\includegraphics[width = 0.45\textwidth]{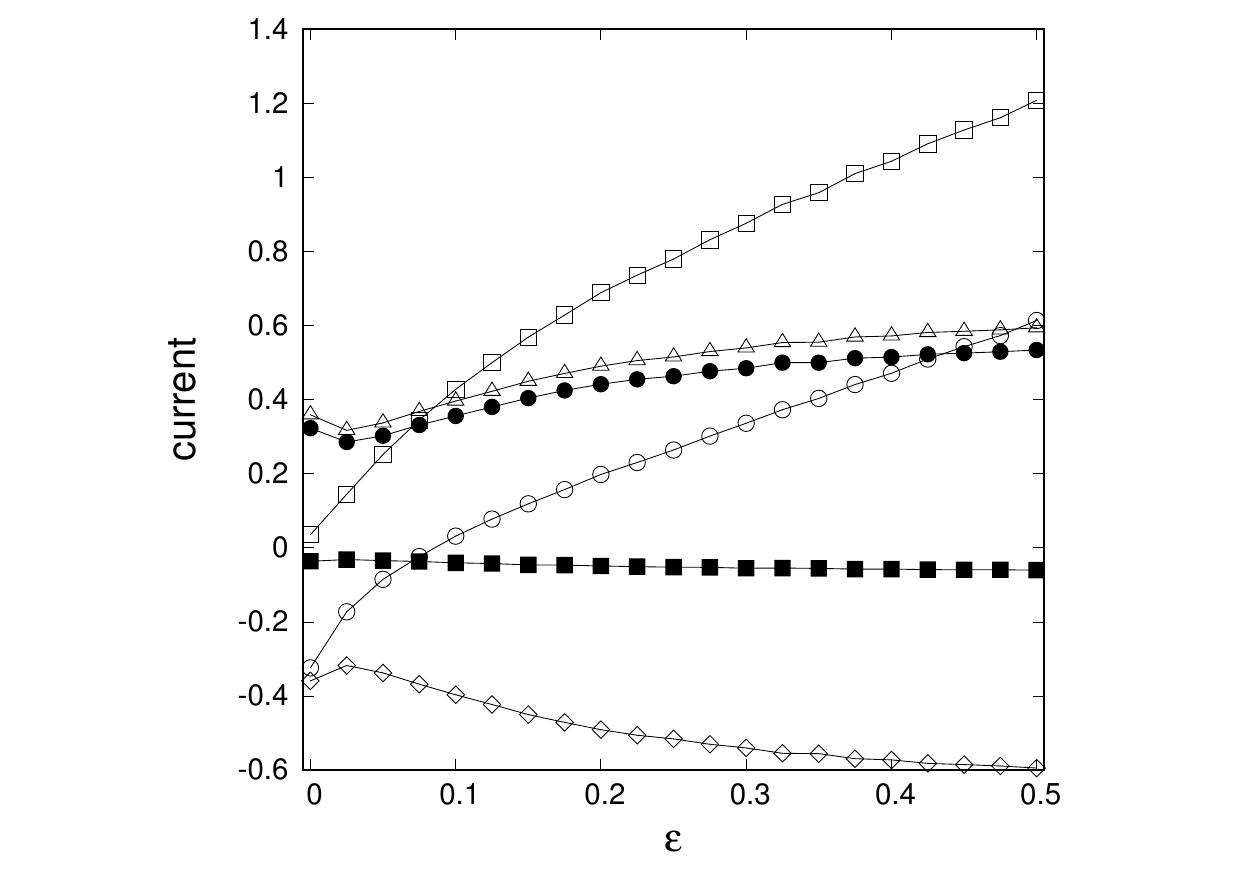} 
\includegraphics[width = 0.45\textwidth]{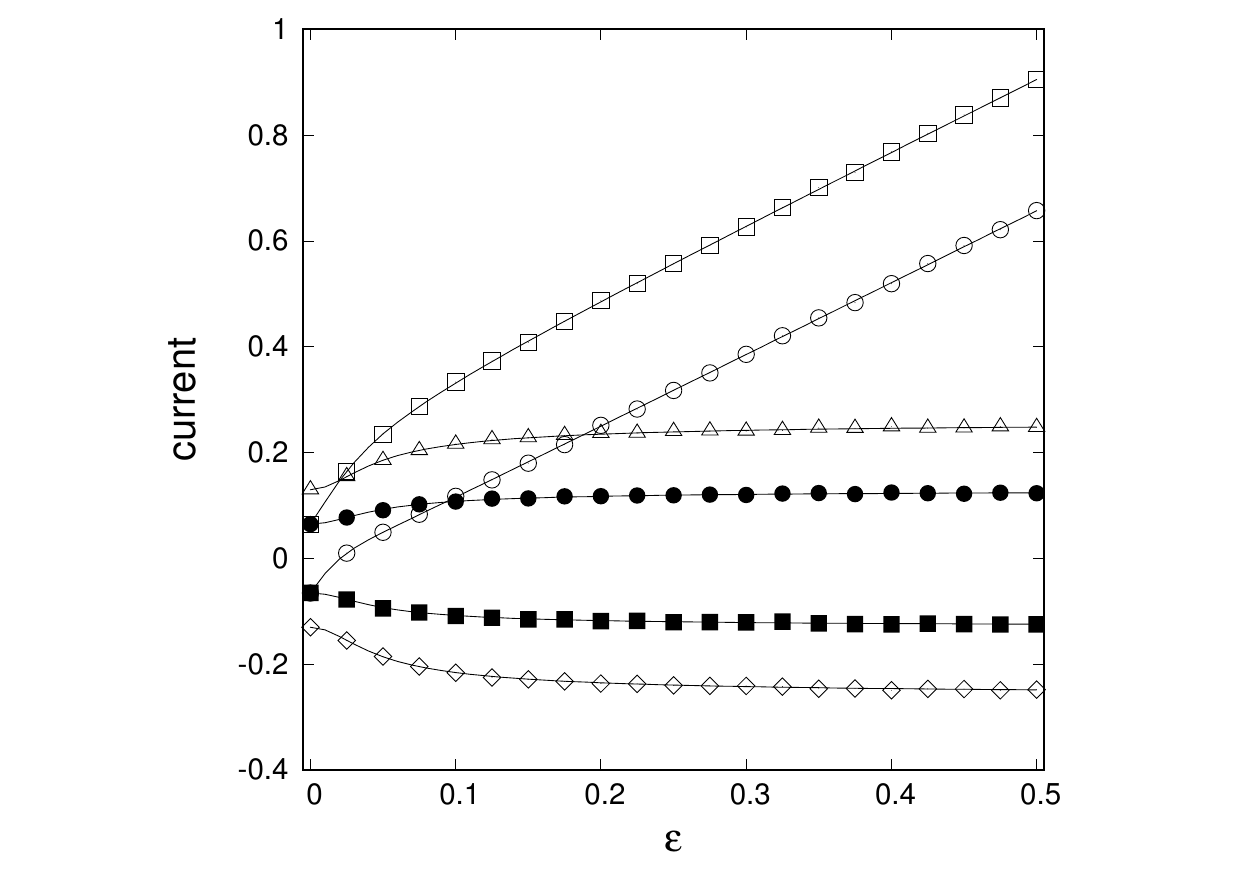} 
\caption{Steady state currents 
$J^\textrm{u}_\Delta$ (empty circles),
$J^\textrm{u}_{L-\Delta}$ (empty squares),
$J^\textrm{d}_\Delta$ (solid circles),
$J^\textrm{d}_{L-\Delta}$ (solid squares),
$J^\textrm{c}_1$ (empty triangles),
and
$J^\textrm{c}_\ell$ (empty diamonds)
as function of the drift $\varepsilon$, 
for 
$\sigma=10$, $\psi=1$, with $L=50$, 
$\Delta=5$ (left panel) and 
$\Delta=L/2$ (right panel). 
Solid lines denote the exact solution.}
\label{fig:fig9}
\end{figure*}

We conclude this subsection discussing the behavior of the stationary currents. 
As remarked in Section~\ref{s:int}, the transient to the stationary 
state is characterized by uphill currents flowing against 
the direction predicted by Fick's law. On the other hand, as we 
show here, the steady--state currents flow as predicted by the Fick's law. 

The steady--state \emph{current}
$J_{i\to j}$ from site $i$ to 
site $j$ is defined as 
the mean value, with respect to the stationary measure, of the quantity 
$n_iW_{j,i}-n_{j}W_{i,j}$
Due to the structure of the jump rates $W_{j,i}$, the 
current $J_{i\to j}$ is nonzero only for neighboring sites $i$ and $j$ 
in the rings and along the channels connecting the two rings. 
Averaging with respect to the stationary measure and 
recalling Eq.~\eqref{ccd470} we have,
\begin{equation}
\label{ccd200}
J_{i\to j}
=
\rho_iW_{j,i}-\rho_{j}W_{i,j}
=
N(
\pi_iW_{j,i}-\pi_{j}W_{i,j})
.
\end{equation}

Thus, at stationarity, we consider the following six currents: 
$J^\textrm{u}_\Delta=J_{L+\ell-1\to L+\ell}$
flows in the branch of the upper ring of length $\Delta$; 
$J^\textrm{u}_{L-\Delta}=J_{L+\ell\to L+\ell+1}$
flows in the branch of the upper ring of length $L-\Delta$ 
from site $L+\ell$ to site $L+1$;
$J^\textrm{d}_\Delta=J_{\ell-1\to \ell}$
flows in the branch of the lower ring of length $\Delta$ 
from site $\ell$ to site $1$;
$J^\textrm{d}_{L-\Delta}=J_{\ell\to \ell+1}$
flows in the branch of the lower ring of length $L-\Delta$ 
from site $\ell$ to site $1$;
$J^\textrm{c}_{1}=J_{L+1\to 1}$
flows in the channel from site $L+1$ to site $1$; and
$J^\textrm{c}_{\ell}=J_{L+\ell\to \ell}$
flows in the channel from site $L+\ell$ to site $\ell$. Note that currents \textit{within} the rings are conventionally taken as positive when particles move counterclockwise on the ring, i.e., 
from a site $x$ to neighbor site $y>x$ (recall that periodic boundary conditions are imposed), whereas currents \textit{between} the rings are taken as positive when flowing from the upper to the lower ring.

The behavior of these currents as functions of the 
drift $\varepsilon$ is shown in Figure~\ref{fig:fig9}: 
on the left we consider a case with small $\Delta$ and on the right 
the case $\Delta=L/2$. Further details are given in the 
caption. 
We first note that, at stationarity, the currents in the channels 
are nonzero, but since they are opposed (empty triangle and empty diamonds),  
the total current between the two rings is zero. 
Another interesting remark is that in the driven ring, the 
currents are approximately  proportional to the drift $\varepsilon$, whereas in the 
symmetric ring the current is nonzero, due to  the 
drift in the upper ring, but approaches a constant value as $\varepsilon$ 
is increased. Finally, it is easy to check that the sign of the 
currents in the undriven ring is consistent with Fick's law when compared with the density profiles plotted in Figures~\ref{fig:fig6} 
and \ref{fig:fig7}.

\subsection{Transient uphill currents}
\label{s:tra}

\begin{figure*}
\includegraphics[width = 0.45\textwidth]{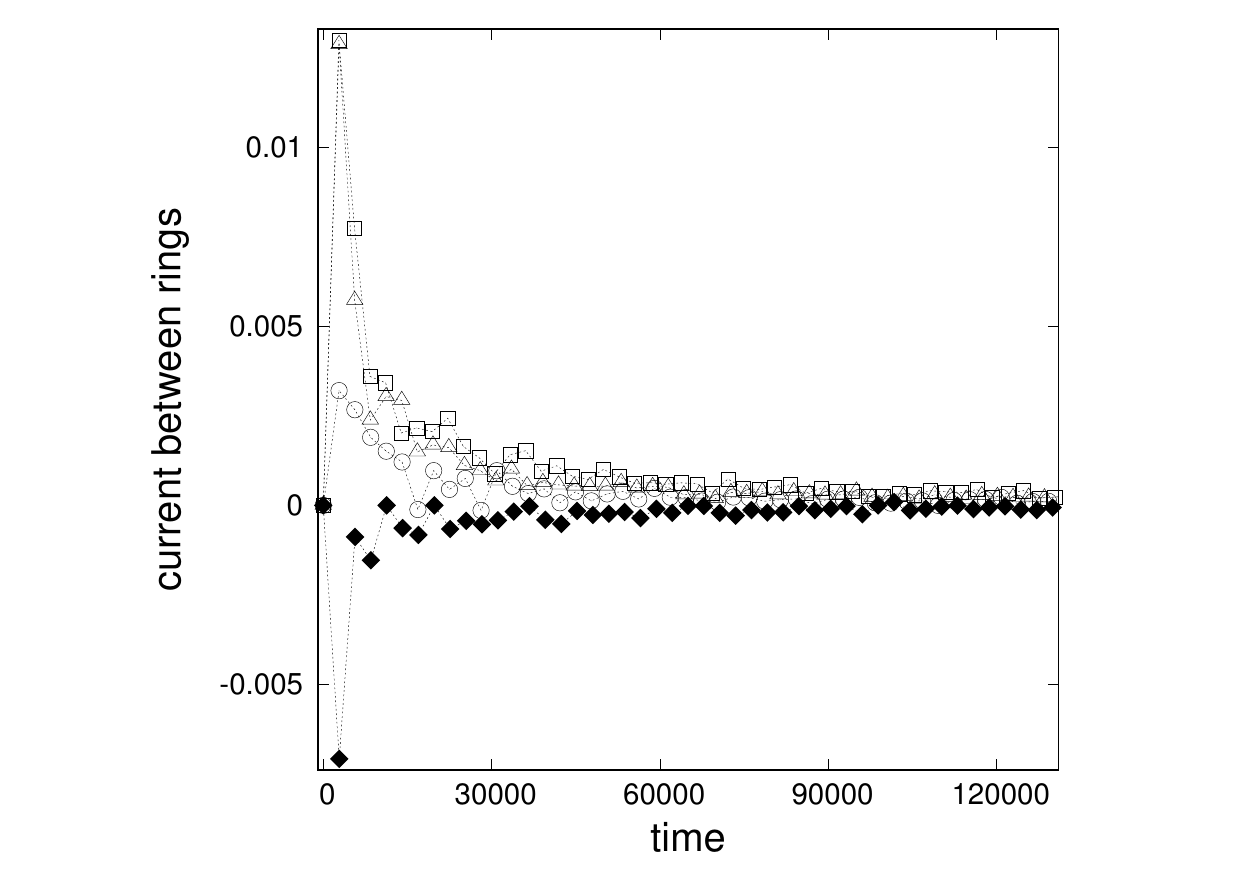} 
\includegraphics[width = 0.45\textwidth]{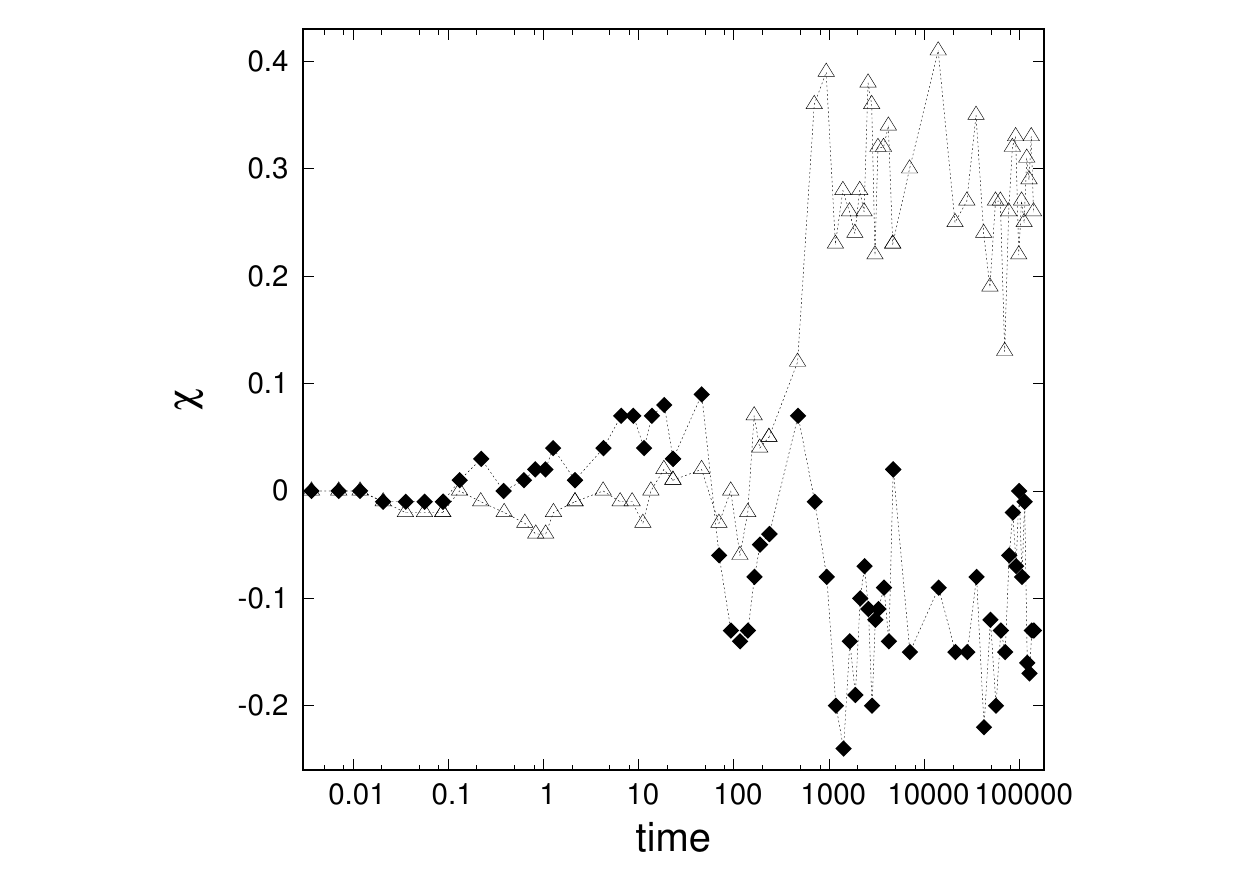} 
\caption{Left: Total current flowing between 
rings (left),
and mass displacement $\chi$ (right),
versus time for $\sigma=10$ and $\psi=1$, 
with $L=50$, $\epsilon=0.05$, and, from the left to the right, 
$\Delta=5$ (solid diamonds), 
$\Delta=15$ (empty circles),
$\Delta=25$ (empty triangles),
$\Delta=35$ (empty squares).
(See the text for further details.)
Dashed lines are a guide for the eye.
On the right, for legibility we plot results for only two values of $\Delta$. 
}
\label{fig:fig10}
\end{figure*}

As a final remark, we observe that the presence of uphill currents is 
highlighted by considering the transient behavior of the total 
current 
flowing between the rings as time goes by.
By the expression \emph{total particle current} at time $t$ we 
refer to the difference between the total number of particles 
which jumped from the upper to the lower ring and those which 
jumped in the opposite direction in the interval from zero to 
$t$, divided by $t$.
(Recall that migration is considered positive when downward).
In the left panel of Figure \ref{fig:fig10}  the total 
current flowing in the channels is shown as a function of time in 
a particular realization of the process for various inter-channel 
distances $\Delta$. The right panel (with logaritmic horizontal scale), 
for the same realization of the process, shows the mass displacement $\chi$ 
as a function of time.

Note that for a moderate drift $\varepsilon=0.05$ and a small 
distance $\Delta=5$ (filled diamonds) the current 
is negative, consistent with the negative average mass displacement $\chi$ 
shown in the right panel.  
The case $\Delta=25$ can be understood similarly.
Note in figure~\ref{fig:fig10}, that when time becomes 
large, the system attains a steady state in which the total current 
exchanged between the rings vanishes tends to zero. Figure \ref{fig:fig9} reveals, 
in particular, that at stationarity $J^\textrm{c}_{1}$ is precisely 
the opposite of $J^\textrm{c}_{\ell}$, 
for any value of the drift $\varepsilon$.

\section{Conclusions}
\label{s:con}

We consider a simple, exactly solvable model of noninteracting particles exhibiting migration despite overall symmetry.  The model consists of independent random walkers hopping on and between a pair of rings,
with particle exchange between rings allowed at specified channels.  The combination
of asymmetry between the channels and a drift in one (but not both) rings leads to particle migration between rings, as shown by the exact solution of the model and confirmed in Monte Carlo simulations.  In the limit of strong channel asymmetry, particle migration
can be understood on the basis of the classic gambler's ruin problem.

Our results show that ``uphill migration'', 
in apparent violation of Fick's law, may be significantly more general than has been appreciated until now, in that
it does not depend on inter--particle interactions.  A promising arena for observation
of this phenomenon is the movement of passive tracers in cyclic geophysical flows.

%{\color{red}  Let me explain the idea.  Think of a cycle like the hydrological cycle, in which %water evaporates over the ocean (channel 1) and condenses and precipitates over %mountains, for example (channel 2).  A rare tracer molecule (like water with the isotope %$^{18}$O) could diffuse at a much higher rate in the atmosphere than in liquid water.  This %might provide the
%ingredients needed for a physical (near) realization of the model. Of course this is quite
%speculative, and we can leave this comment out if you prefer.  On the other hand, I know
%a climate physicist who might be interested in this idea.}

\vskip 1 cm 

%Per i ringraziamenti
\begin{acknowledgments}
MC acknowledges A.\ De Masi and E.\ Presutti for useful discussions.
ENMC acknowledges A.\ Ciallella for useful discussions.
RD acknowledges support from CNPq, Brazil, through project 
number 303766/2016--6.
ENMC and MC acknowledge support from FFABR2017.
The authors thank an anonymous referee whose comments helped them 
to improve the article. 
\end{acknowledgments}

%Eventuali appendici
%\appendix
%\renewcommand{\theequation}{\Alph{section}.\arabic{equation}}

%%%%%%%% Bibliografia
%\newpage
%\addcontentsline{toc}{section}{References}
%\newpage
%\begin{thebibliography}{ZZZZZ}


\begin{thebibliography}{99}

\bibitem{D14}
R.\ Dickman, 
\textit{Failure of steady--state thermodynamics in nonuniform driven 
lattice gases}, 
2014, Phys.\ Rev.\ E \textbf{90}, 062123.

\bibitem{guioth2018}
J. Guioth and E. Bertin,
\textit{Large deviations and chemical potential in bulk--driven 
systems in contact},
2018, Europhys.\ Lett.\ \textbf{123}, 10002.

\bibitem{CDMP16} M.~Colangeli, A.~De~Masi, E.~Presutti,
\textit{Latent heat and the Fourier law},
2016, Phys.\ Lett.\ A \textbf{380}, 1710--1713.

\bibitem{CDMP17} M.~Colangeli, A.~De~Masi, E.~Presutti, 
\textit{Particle models with self sustained current},
2017, J.\ Stat.\ Phys.\ \textbf{167}, 1081--1111.

\bibitem{CGGV18}
M.~Colangeli, C.~Giardin\`{a}, C.~Giberti, C.~Vernia, 
\textit{Nonequilibrium two-dimensional Ising model with stationary uphill diffusion},
2018, Phys.\ Rev.\ E \textbf{97}, 030103(R).

\bibitem{ladder07}
R.~Dickman and R.~R.~Vidigal, 
\textit{Particle redistribution and slow decay of correlations in 
hard--core fluids on a half-driven ladder}, 
2007, J.\ Stat.\ Mech.\ P05003.

\bibitem{EH05}
M.R.\ Evans, T.\ Hanney, 
\textit{Nonequilibrium statistical mechanics of the zero--range 
process and related models},
2005, J.\ Phys.\ A: Math.\ Gen.\ \textbf{38}, R195.

\bibitem{Ligget}
T.H.~Ligget,
Continuous Time Markov Processes,
An Introduction,
Graduate Studies in Mathematics, vol.\ 113, 
American Mathematical Society, Providence, Rhode Island, 2010.

\bibitem{CC17}
E.N.M.\ Cirillo, M.\ Colangeli, 
\textit{Stationary uphill currents in locally perturbed zero--range processes},
2017, Phys.\ Rev.\ E \textbf{96}, 052137.

\bibitem{CKMSpre2016}
E.N.M.\ Cirillo, O.\ Krehel, A.\ Muntean, and R.~van Santen,
\textit{Lattice model of reduced jamming by a barrier},
2016, Phys.\ Rev.\ E \textbf{94}, 042115.

\bibitem{CCSpre2018}
A.\ Ciallella, E.N.M.\ Cirillo, J.\ Sohier.
\textit{Residence time of symmetric random walkers in a strip with 
large reflective obstacles},
2018, Physical Review E \textbf{97}, 052116.

\bibitem{Feller1968}
W.\ Feller.
An Introduction to Probability Theory and its Applications,
volume~1.
John wiley \& Sons, Inc, New York -- London -- Sidney, 1968.

\bibitem{CCpreprint2018}
A.\ Ciallella and E.N.M. Cirillo,
\textit{Conditional expectation of the duration of the classical gambler 
problem with defects},
in press on Eur.\ Phys.\ J.\ Spec.\ Top..

\end{thebibliography}
\end{document}